Moisture-induced superconductivity in FeTe$_{0.8}$S$_{0.2}$

Y. Mizuguchi[1,2,3], K. Deguchi[1,2,3], S. Tsuda[1,2], T. Yamaguchi[1,2] and Y. Takano[1,2,3]
1. National Institute for Materials Science, 1-2-1 Sengen, Tsukuba, 305-0047, Japan
2. Japan Science and Technology Agency-Transformative Research-Project on Iron-Pnictides, 1-2-1 Sengen, Tsukuba, 305-0047, Japan
3. Graduate school of Pure and Applied Sciences, University of Tsukuba, 1-1-1 Tennodai, Tsukuba, 305-8571, Japan

Abstract

Moisture-induced superconductivity was observed in FeTe$_{0.8}$S$_{0.2}$. With exposing the sample to the air, the zero resistivity temperature and the superconducting volume fraction were enhanced up to 7.2 K and 48.5 %, respectively, while the as-grown sample showed only filamentary superconductivity. We concluded that the causes of the evolution of superconductivity were water-related ions and/or molecules, because only the sample kept in water at room temperature for several days showed superconductivity. The speed of evolution of superconductivity was strongly enhanced by immersing the sample into the hot water.

I. Introduction

Iron chalcogenides attract researchers as the simplest iron-based superconductors.[1,2] The tetragonal FeSe superconductor shows dramatic pressure effect on transition temperature $T_c$; the onset temperature $T_c^{onset}$ increases from 13 to 37 K at 4-6 GPa.[3-6] Crystal structural analysis and NMR studies under high pressure indicated that the enhancement of $T_c$ was related to the change in the crystal structure or the magnetic states.[4,5,7] Correlation between superconductivity and magnetism is likely to be important to understand the mechanism of superconductivity in the iron chalcogenides. In fact, tetragonal FeTe, which has a structure analogous to superconducting FeSe, undergoes antiferromagnetic ordering at ~70 K and does not show superconductivity. The magnetic ordering is suppressed by S or Se substitution, and superconductivity appears.[8-11] However, hydrostatic pressure could not induce superconductivity in FeTe.[12,13] To clarify the reason why only the Te-site substitution can induce superconductivity in FeTe, we focus on S-substituted FeTe.

An optimum way to synthesize the high-quality superconducting sample of $FeTe_{1-x}S_x$ has not been established yet, probably due to a solubility limit of S for the Te site. The superconducting properties of $FeTe_{1-x}S_x$ depended on the sample preparation method.[8] The $FeTe_{0.8}S_{0.2}$ sample synthesized by the melting method showed a sharp superconducting transition at $T_c^{onset}$ = 10.5 K; however, the obtained sample contained impurity phases. On the other hand, the $FeTe_{0.8}S_{0.2}$ sample synthesized using the solid-state reaction method was almost the single phase; however the solid-state-reacted sample showed a broad transition in temperature dependence of resistivity, and diamagnetism corresponding to superconductivity was not observed. The cause of filamentary superconductivity in the solid-state-reacted sample would be an insufficiency of the shrinkage of lattice, in other words, an insufficiency of S concentration. Here we report moisture-induced superconductivity in $FeTe_{0.8}S_{0.2}$ synthesized by the solid-state reaction.

II. Experimental details

The polycrystalline samples of $FeTe_{0.8}S_{0.2}$ were prepared using the solid-state reaction method as described in Ref. 8. At first, we synthesized the TeS precursor by reacting the Te (99.9 %) and S (99 %) powders to produce a homogeneous reaction. The Te and S powders were sealed into an evacuated quartz tube, heated at 500 °C for 10 h, and furnace-cooled. Then the powders of Fe (99.9 %), Te (99.9 %) and TeS were sealed into an evacuate quartz tube with a nominal composition of $FeTe_{0.8}S_{0.2}$, and heated at 600 °C for 15 h. After furnace cooling, the products were ground, palletized, sealed into the evacuated quartz tube and heated again at 600 °C for 15 h. Temperature dependence of resistivity was measured down to 2 K using the four-terminals method. Temperature dependence of susceptibility after both zero field cooling (ZFC) and field cooling (FC) was measured using a SQUID magnetometer down to 2 K under a magnetic field of 10 Oe. Powder x-ray diffraction patterns were collected using the Cu-Kα radiation. The room temperature of the laboratory was kept at 20 ~ 25 ºC.

III. Results and discussion

Figure 1 shows the temperature dependence of resistivity for $FeTe_{0.8}S_{0.2}$ with several air-exposure time from 0 to 110 days. For the as-grown sample, zero resistivity was not observed while an onset of the superconducting transition was observed at 8.0 K. The diamagnetic signal was not observed in the susceptibility measurement, indicating the absence of bulk superconductivity. Surprisingly, after exposing the sample to the air for 2 days, zero resistivity appeared around $T_c^{zero}$ = 2 K. With increasing air-exposure time, both the $T_c^{onset}$ and $T_c^{zero}$ increased up to 10.2 K and 7.2 K, respectively. The superconducting transition became sharper with increasing air-exposure time. Figure 2 shows the temperature dependence of resistivity for as-grown $FeTe_{0.8}S_{0.2}$, 110-day-old $FeTe_{0.8}S_{0.2}$ and $Fe_{1.08}Te$. For $Fe_{1.08}Te$, we can find an anomaly corresponding to the structural and magnetic transition around 70 K. The anomaly seems to be suppressed completely for as-grown $FeTe_{0.8}S_{0.2}$; however, bulk superconductivity was not observed. After 110 days, the sharp superconducting transition appeared. The normal-state resistivity of 110-day-old $FeTe_{0.8}S_{0.2}$ is clearly lower than that as grown, implying the change in the carrier density by the air exposure.

Figure 3 shows the temperature dependence of magnetic susceptibility for $FeTe_{0.8}S_{0.2}$ with several air-exposure time from 20 to 140 days. Although we could not observe the superconducting transition for the as-grown sample, the diamagnetic signal corresponding to superconductivity appeared for 20-day-old $FeTe_{0.8}S_{0.2}$. With increasing air-exposure time, the $T_c$ increased and the diamagnetic signal was enhanced. The $T_c$ estimated from susceptibility ($T_c^{mag}$) was plotted in Fig. 4 as a function of air-exposure time with the $T_c^{onset}$ and $T_c^{zero}$ determined from the resistivity measurements. The $T_c^{mag}$ almost corresponded to the $T_c^{zero}$ and reached 7.2 K after 140 days. The superconducting volume fraction was calculated from a difference between the value of the normal state and 2 K, and plotted in Fig. 4 as a function of air-exposure time. The superconducting volume fraction was dramatically enhanced up to 48.5 %, indicating that the bulk superconductivity was induced by the air exposure while as-grown $FeTe_{0.8}S_{0.2}$ showed only filamentary superconductivity.

To clarify the origin of the dramatic change in the superconducting properties induced by the air exposure, we carried out the powder x-ray diffraction for FeTe$_{0.8}$S$_{0.2}$ just after the synthesis and after 100 and 200 days. Figure 5(a) shows the x-ray diffraction patterns for as-grown and 200-day-old FeTe$_{0.8}$S$_{0.2}$. The peaks were indexed using the *P*4/*nmm* space group. Lattice constants were calculated to be $a$ = 3.8158(9) and $c$ = 6.2445(23) Å for as-grown FeTe$_{0.8}$S$_{0.2}$, $a$ = 3.8114(8) and $c$ = 6.2421(21) Å for 100-day-old FeTe$_{0.8}$S$_{0.2}$, and $a$ = 3.8097(8) and $c$ = 6.2307(20) Å for 200-day-old FeTe$_{0.8}$S$_{0.2}$, respectively. The calculated lattice constants $a$ and $c$ were plotted in Fig. 5(b) and (c) as a function of air-exposure time. The lattice constants slightly decreased with the air exposure for 200 days. The shrinkage of lattice should be related to the dramatic change in the superconducting properties induced by air exposure.

To investigate the factor that induced superconductivity, we measured temperature dependence of susceptibility for the samples kept in several conditions. The as-grown FeTe$_{0.8}$S$_{0.2}$ samples were kept in vacuum (<0.5 Pa), ion-exchanged water, O$_2$ gas and N$_2$ gas for several days at room temperature. Figure 6(a), (b) and (c) show the typical temperature dependence of susceptibility for the samples kept in vacuum for 40 days, water for 10 days and O$_2$ gas for 40 days, respectively. The superconducting transition was observed only for the sample kept in water, although the samples kept in both vacuum and O$_2$ gas did not show the superconducting transition. Also the sample kept in N$_2$ gas for 40 days did not show the superconducting transition. Therefore we concluded that the moisture in the air induced the dramatic change in the superconducting properties. We also investigated the magnetic properties of Fe$_{1.08}$Te kept in the water for 50 days as shown in Fig. 6(d). There was no sign of superconductivity. In fact, moisture-induced superconductivity is unique for FeTe$_{1-x}$S$_x$ among the iron chalcogenides.

Recently, water-induced superconductivity was reported also in SrFe$_2$As$_2$, which is one of the parent phases of the iron-based superconductors.[14] The authors suggested that superconductivity was induced when the lattice was compressed by exposing the sample to H$_2$O-related species. Moisture-induced superconductivity in FeTe$_{0.8}$S$_{0.2}$ might be induced as in SrFe$_2$As$_2$. In the case of the water-intercalated superconductor Na$_x$CoO$_2 \cdot y$H$_2$O, superconductivity appeared when the *c* axis was expanded by a partial substitution of Na$^+$ ion by H$_3$O$^+$ ion.[15] On the basis of the shrinkage of lattice in the air-exposed FeTe$_{0.8}$S$_{0.2}$, the H$_3$O$^+$ ion would not be the origin of moisture-induced superconductivity.

In this respect, one of the candidate elements to explain moisture-induced superconductivity in FeTe$_{0.8}$S$_{0.2}$ is the H$^+$ ion because the ionic radius is very small.

Similar situation was reported in the $Li^+$-intercalated $KCa_2Nb_3O_{10}$ superconductor. Superconductivity was induced by the $Li^+$ intercalation without any change in the lattice constants.[16] The $H^+$-ion intercalation into the interlayer of $FeTe_{0.8}S_{0.2}$ would not change the lattice constants largely. If the $H^+$ was intercalated, the electron carriers should be generated in the Fe layer. The decrease of resistivity for the air-exposed $FeTe_{0.8}S_{0.2}$ as shown in Fig. 2 would be originated in the increase of the electron carrier density. It might completely suppress the magnetism that had barely survived in as-grown $FeTe_{0.8}S_{0.2}$. Another candidate is the $OH^-$ ion. If the $OH^-$ ion were intercalated to the interlayer site, it would compress the lattice because there is the excess Fe at the interlayer site.

The $O^{2-}$ ion is also a candidate to explain this phenomenon, because a solution of oxygen in the water should not be neglected. Furthermore, $O^{2-}$ has a valence same with $S^{2-}$ and $Te^{2-}$, and its ionic radius is the smallest among them. The shrinkage of lattice might be due to the partial substitution of O for the Te/S site. However, the sample kept in $O_2$ gas for 40 days did not show superconductivity. If the origin of moisture-induced superconductivity in $FeTe_{0.8}S_{0.2}$ were the $O^{2-}$ ion, the water should play an important role to induce superconductivity as a catalyst at room temperature. In any case, the microscopic investigations sensitive to a local structure such as EXAFS (extended x-ray absorption fine structure) are needed to determine the element that induced superconductivity.

If any chemical intercalation or substitution would occur in the moisture, the speed of the reaction should depend on the surrounding temperature. In this respect, we investigated the temperature dependence of susceptibility for the samples kept in the hot water (ion-exchanged water) with a temperature of ~70 ºC. Figure 7 shows the temperature dependence of normalized susceptibility for the two samples kept in the hot water for 4 and 24 hours. Although the sample kept in the cold water (at room temperature) for a few days showed no sign of superconductivity, the superconducting transition was observed for the sample kept in the hot water for only 4 hours. Furthermore, with increasing immersion time, both $T_c$ and the superconducting volume fraction were enhanced up to 6.8 K and 15.6 %, respectively. This suggests that the phenomena that we reported here are much sensitive to the surrounding temperature.

IV. Conclusion

We reported moisture-induced superconductivity in $FeTe_{0.8}S_{0.2}$ synthesized by the solid-state reaction method. With increasing air-exposure time, the $T_c$ and superconducting volume fraction were enhanced up to 7.2 K and 48.5 %, respectively, while the as-grown sample showed only filamentary superconductivity. The shrinkage of lattice was observed by the air exposure. We concluded that the origin of the dramatic change in superconductivity was moisture in the air, because only the sample kept in water at room temperature showed superconductivity. The speed of evolution of superconductivity was strongly enhanced by immersing the sample into the hot water, indicating the sensitivity of moisture-induced superconductivity to the surrounding temperature.

Acknowledgement
　This work was partly supported by Grant-in-Aid for Scientific Research (KAKENHI).

Figure captions

FIG. 1. Temperature dependence of resistivity for the FeTe$_{0.8}$S$_{0.2}$ sample kept in the air for several days.

FIG. 2. Temperature dependence of resistivity for as-gown FeTe$_{0.8}$S$_{0.2}$, 110-day-old FeTe$_{0.8}$S$_{0.2}$ and Fe$_{1.08}$Te. The anomaly observed in Fe$_{1.08}$Te around 70 K was not observed for both as-gown FeTe$_{0.8}$S$_{0.2}$ and 110-day-old FeTe$_{0.8}$S$_{0.2}$, indicating that the magnetic ordering was suppressed by the S substitution.

FIG. 3. Temperature dependence of magnetic susceptibility for the FeTe$_{0.8}$S$_{0.2}$ sample kept in the air for several days.

FIG. 4. Air-exposure time dependence of the $T_c^{onset}$, $T_c^{zero}$, $T_c^{mag}$ and the superconducting volume fraction. The horizontal axis is logarithmic.

FIG. 5. (a) Powder x-ray diffractin patterns collected just after the synthesis and after 200 days. All peaks of the tetragonal FeTe$_{0.8}$S$_{0.2}$ phase were indexed using the *P4/nmm* space group. The astarisks indicate the impurity phases. The calculated lattice constants *a* and *c* were plotted in (b) and (c) as a function of air-exposure time, respectively.

FIG. 6. Temperature dependence of magnetic susceptibility for (a)FeTe$_{0.8}$S$_{0.2}$ kept in vacuum, (b)FeTe$_{0.8}$S$_{0.2}$ kept in water, (c)FeTe$_{0.8}$S$_{0.2}$ kept in O$_2$ gas, and (d)Fe$_{1.08}$Te kept in water. The superconducting transition was observed only in (b).

FIG. 7. Temperature dependence of normalized susceptibility for the samples kept in the hot water for 4 and 24 hours, respectively.

Fig. 1

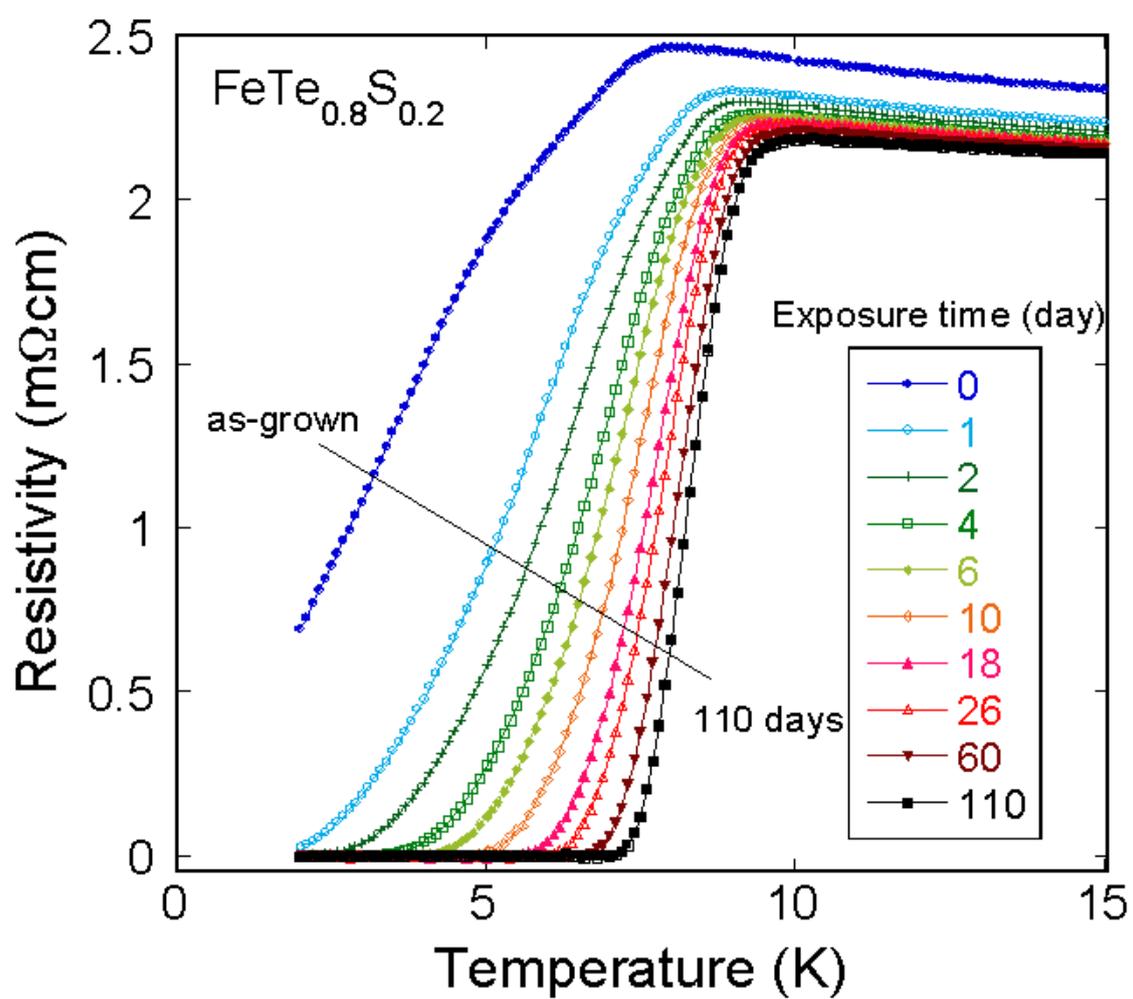

Fig. 2

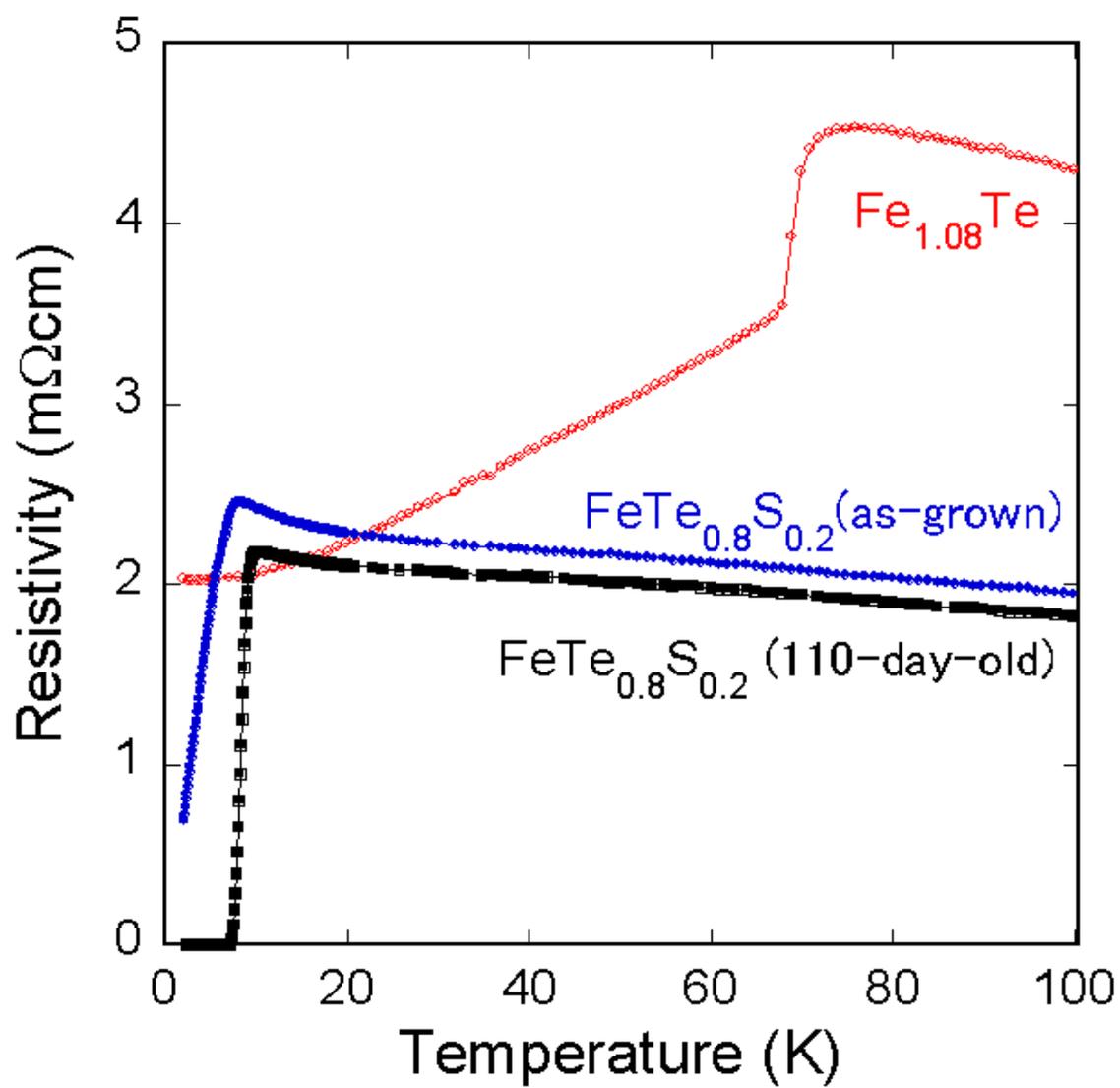

Fig. 3

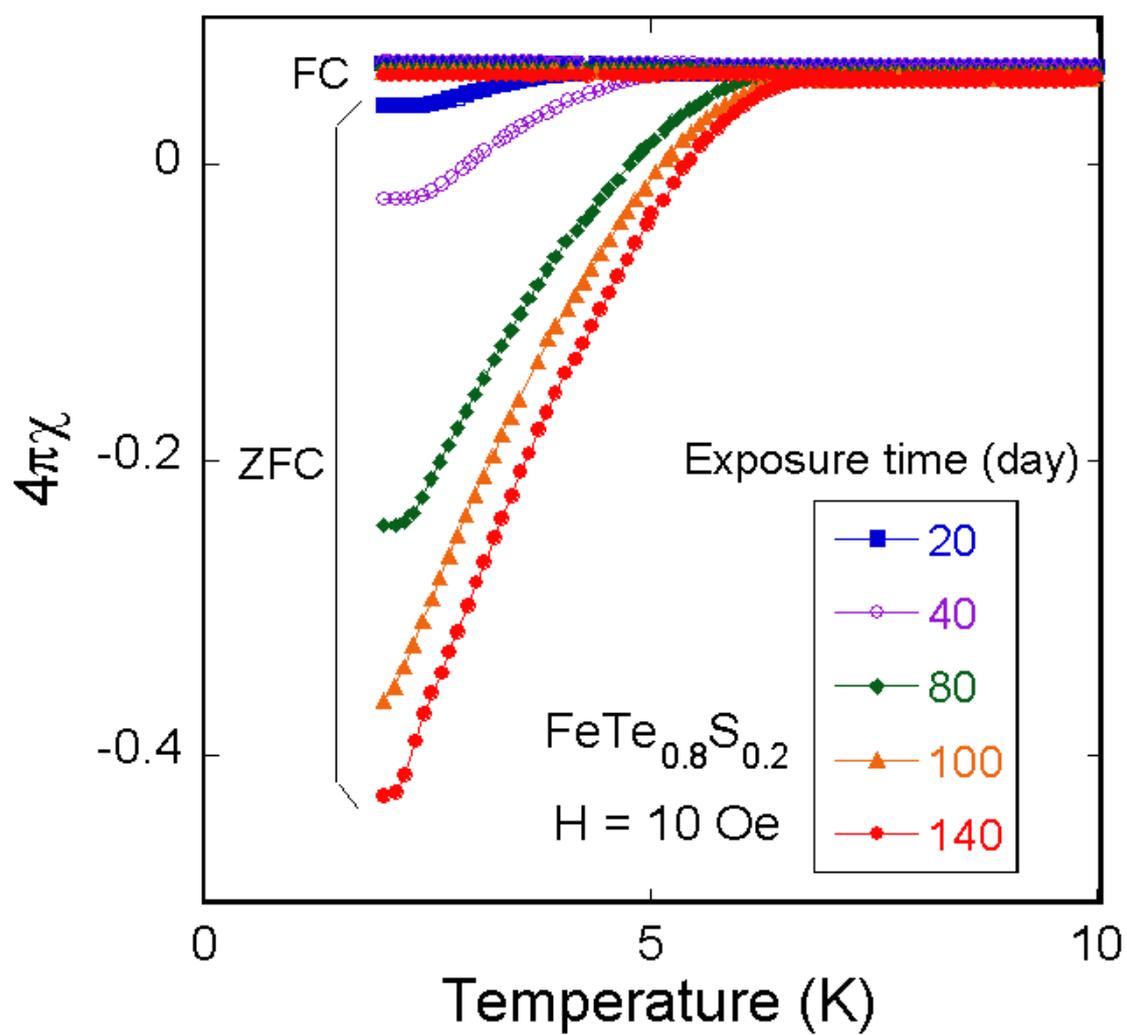

Fig. 4

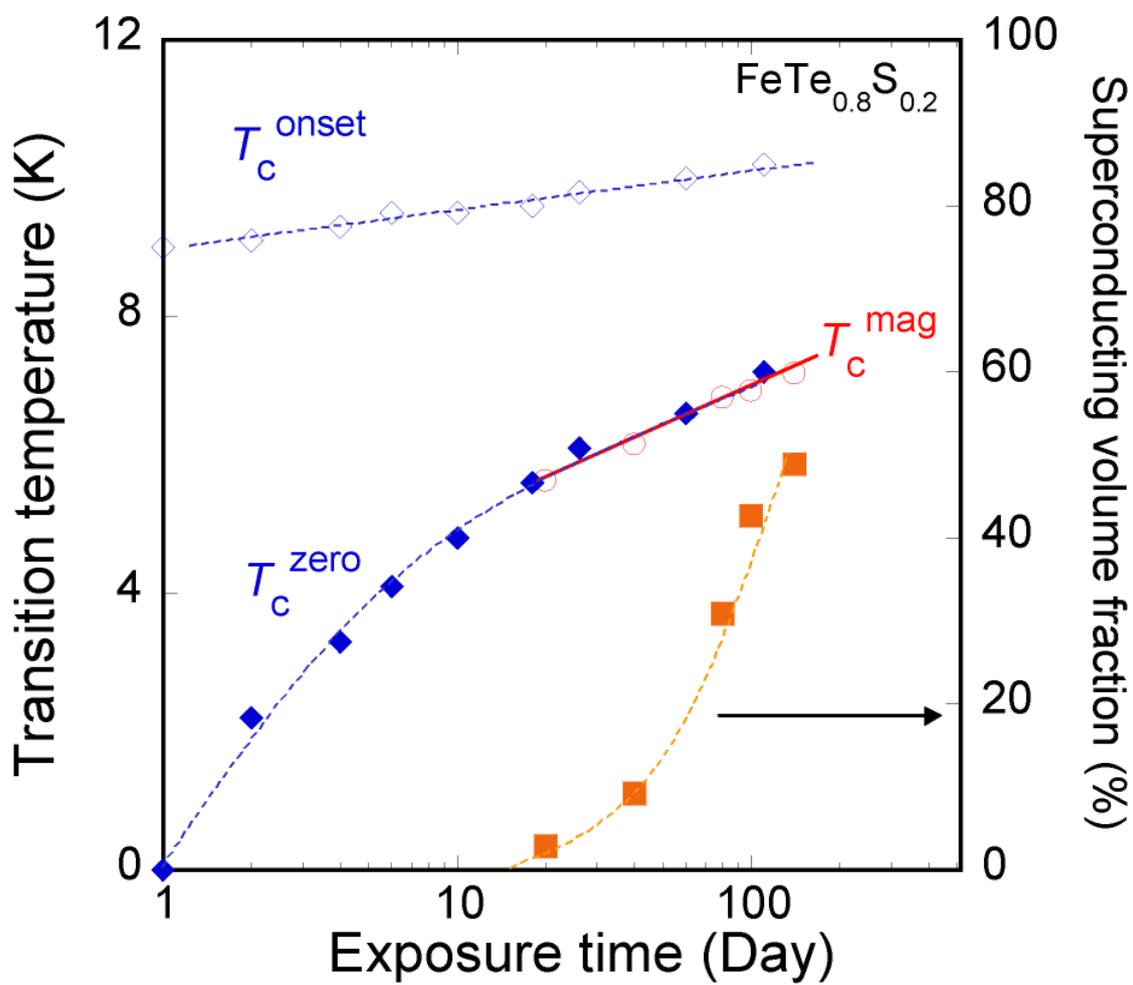

Fig. 5

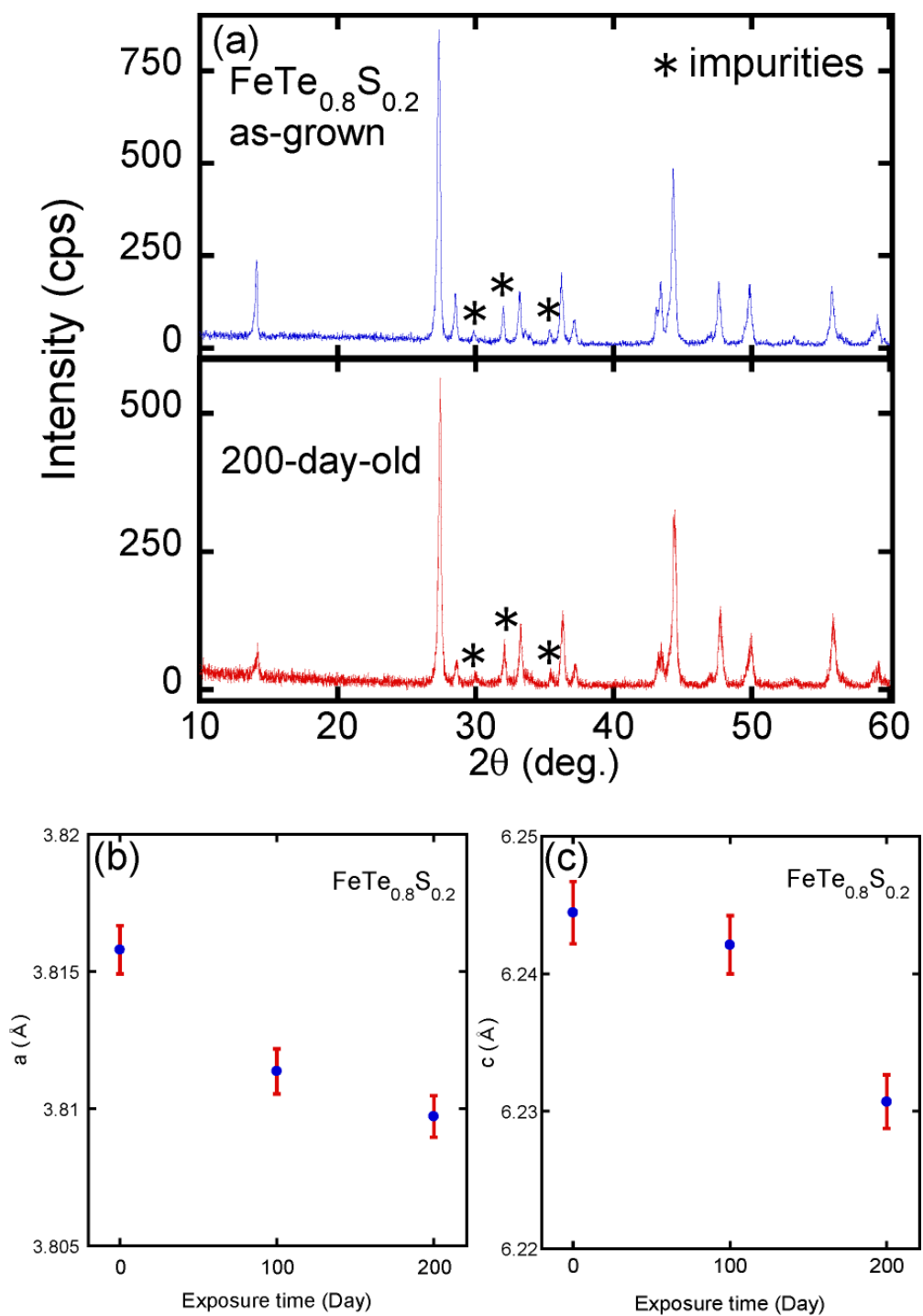

Fig. 6

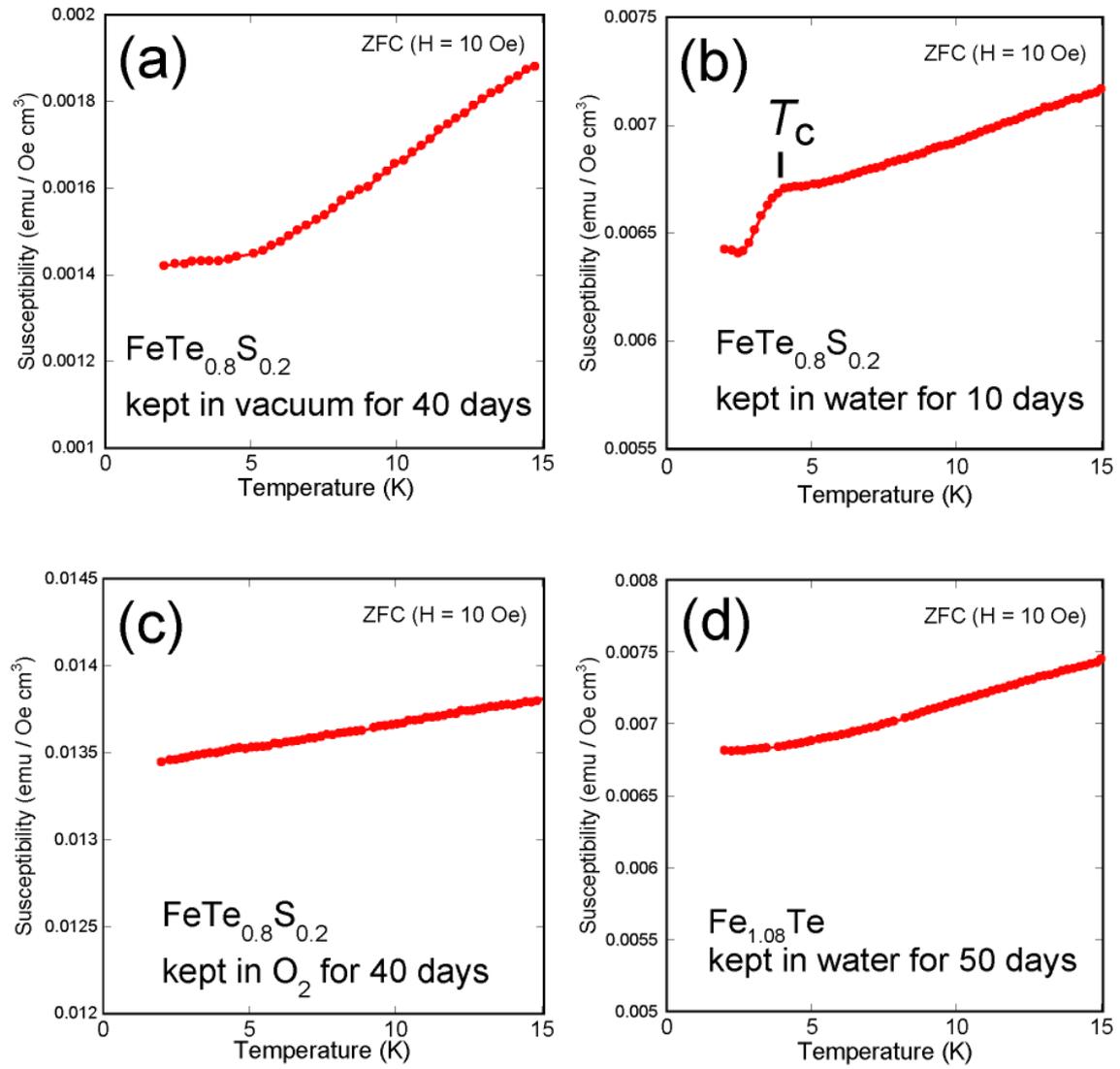

Fig. 7

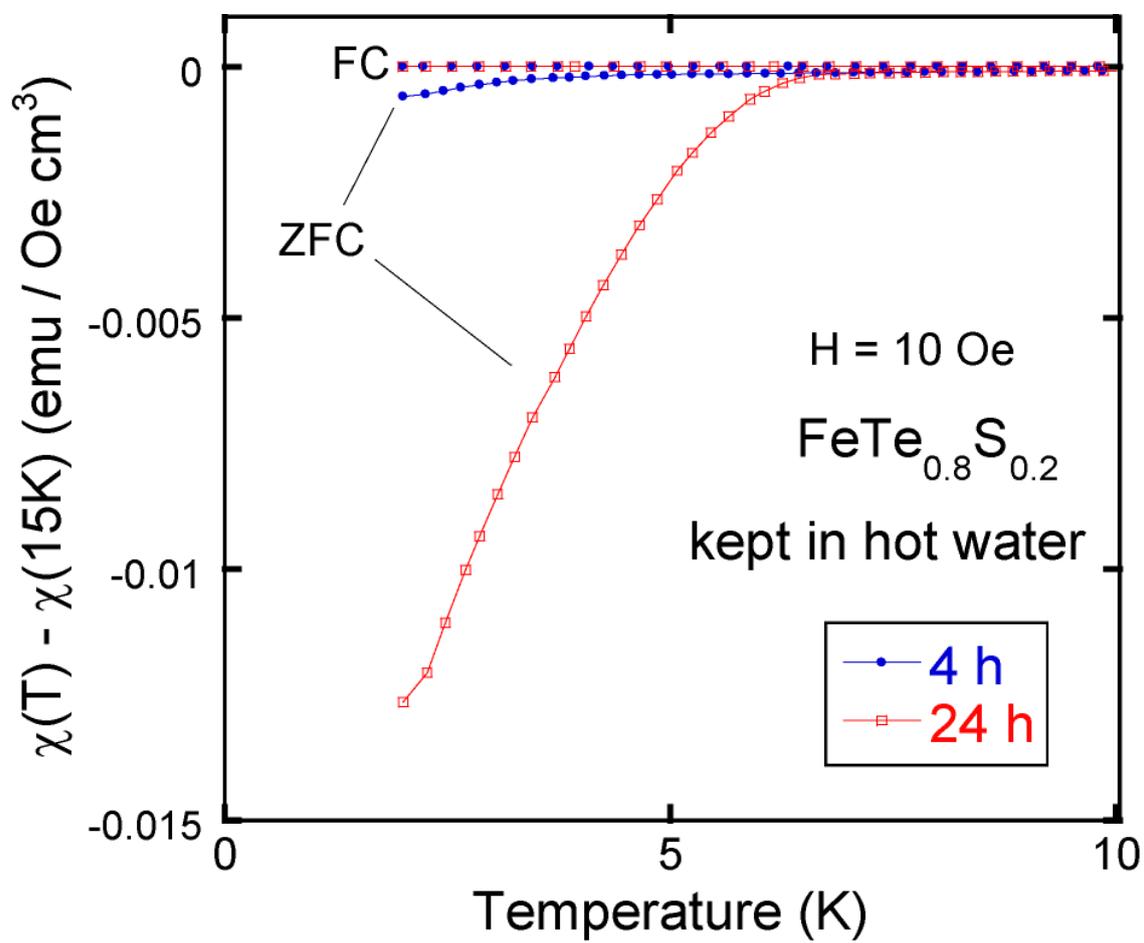